\documentclass[final,1p,times,twocolumn,numbers]{elsarticle}
\usepackage{amssymb}
\usepackage{amsmath}
\usepackage{lineno}

\usepackage{comment}

\usepackage{siunitx}
\usepackage{textcomp}
\usepackage{xcolor}  % make sure this is included
\usepackage{xspace}
\usepackage[version=4]{mhchem}%
\usepackage{url}
\usepackage{hyperref}
\makeatletter
\renewcommand{\url}[1]{} % removes the URL itself
\makeatother

\journal{Journal of Computational Physics}

\begin{document}

\begin{frontmatter}

\title{g4chargeit: Geant4-based kinetic Monte Carlo simulations of charging in dielectric materials}

\author[1,2]{Kush P. Gandhi}
\author[1,3]{Advik D. Vira}
\author[4]{William M. Farrell}
\author[1]{Nikolai Simonov}
\author[5]{Alvaro Romero-Calvo}
\author[1,6]{Thomas M. Orlando}
\author[1]{Phillip N. First}
\author[1]{Zhigang Jiang}

\affiliation[1]{organization={School of Physics, Georgia Institute of Technology}, city={Atlanta}, state={GA}, country={USA}}
\affiliation[2]{organization={Department of Physics, Duke University}, city={Durham}, state={NC}, country={USA}}
\affiliation[3]{organization={Space Science and Applications, Los Alamos National Laboratory}, city={Los Alamos}, state={NM}, country={USA}}
\affiliation[4]{organization={Space Science Institute}, city={Boulder}, state={CO}, country={USA}}
\affiliation[5]{organization={Daniel Guggenheim School of Aerospace Engineering, Georgia Institute of Technology}, city={Atlanta}, state={GA}, country={USA}}
\affiliation[6]{organization={School of Chemistry and Biochemistry, Georgia Institute of Technology}, city={Atlanta}, state={GA}, country={USA}}

% Corresponding author mailing address and e-mail address:
%\correspondingauthor{Advik D. Vira}{avira@gatech.edu}

%% Abstract
\begin{abstract}
We present \texttt{g4chargeit}, a kinetic Monte Carlo framework built on Geant4 for simulation of time-dependent electrostatic charging in dielectric materials. The model explicitly incorporates stochastic particle transport and scattering processes using validated Geant4 cross-sections, while self-consistently evolving the electric potential and field. As a representative application, we simulate the charging of lunar regolith grains under average dayside conditions. The surface of the Moon, in addition to other airless planetary bodies, is regularly exposed to solar ultraviolet photons and solar-wind plasma, creating a radiation environment in which electrostatic interactions among regolith grains become significant. Until now, simulations of regolith charging have often relied on analytical approximations that oversimplify grain geometry and interaction mechanisms. Our Geant4-based simulations reveal charge accumulation within intergrain micro-cavities, leading to repulsive electrostatic forces consistent with experimental observations. The framework establishes a multiscale approach that links microscopic scattering events to the continuity equation of surface charge density and to the formation of macroscopic surface charge patches in complex grain geometries. Although demonstrated here for planetary regolith, the method is general and applicable to a broad range of dielectric charging problems. The code is openly available at \href{https://github.com/kgandhi63/g4chargeit.git}{https://github.com/kgandhi63/g4chargeit.git}.
\end{abstract}

%% Keywords

\end{frontmatter}



%\begin{linenumbers}
\section{Introduction}\label{sec:intro}

Kinetic Monte Carlo (KMC) simulations of dielectric materials are of broad interest across multiple fields, including the modeling of photovoltaic devices \cite{Casalegno2010}, electrostatic forces in DNA packing \cite{Guldbrand1986, Sun2021}, radiation transport in semiconductors \cite{Wang1985, Akturk2007}, and dielectric breakdown in space equipment \cite{Yu2016,wang2022_geant4comsol}. Radiation transport is commonly simulated using Monte Carlo (MC) codes because they provide a general and computationally efficient framework for sampling the microscopic scattering, energy-loss, and secondary-particle processes that govern particle trajectories in matter. These stochastic collision histories are localized transport events, but together they determine how charge is deposited in space and time, driving macroscopic electrostatic fields, current densities, and charge evolution in matter. Conventional MC radiation-transport codes often lack a native mechanism for self-consistently modeling time-evolving phenomena, such as the accumulation of electrostatic fields and their feedback on subsequent particle motion \cite{Akturk2007,Fichthorn_1991}. As a result, charge-accumulation problems frequently require multiple interdependent simulation toolkits, coupling MC-based transport statistics to separate electrostatic or field solvers and leading to hybridized computational models \cite{Yu2016,wang2022_geant4comsol,Wu2025,Ramachandran2025}.

Several computational approaches have been developed to address related charging problems in dusty plasmas and plasma-material interfaces. Early models of dusty plasmas around spacecraft used particle-in-cell Monte Carlo collision (PIC-MCC) methods, in which plasma species and dust grains are advanced statistically while the electric field is obtained from Poisson's equation \cite{Conger1992,Erlandson1994}. Other fully kinetic approaches, such as particle-particle particle-mesh (P3M) models, have resolved plasma trajectories near individual dust grains to capture particle absorption, dust charging, and ion-drag forces \cite{Matyash2010}. Adaptive-mesh-refinement particle-in-cell (AMR-PIC) methods have also been developed to couple particle-based transport with dynamically refined mesh solutions, improving resolution and scalability for multiscale particle-field problems \cite{Gassmoller2018}. Additionally, immersed-finite-element PIC methods have enabled self-consistent simulations of dielectric surface charging by resolving plasma particles, dielectric interfaces, surface-charge deposition, and electrostatic fields on mesh-based domains \cite{Han2016,Han2021,Lund2024}. These PIC-based and adaptive mesh approaches are powerful for resolving plasma sheath dynamics, orbital-motion-limited charging, multiscale transport, and field distributions around dielectric bodies. However, they are typically formulated to address charging at plasma-material interfaces, whereas radiation-driven charging problems require detailed treatment of stochastic material-interaction processes, including photon, electron, and ion transport, secondary-particle generation, charge deposition, and the subsequent evolution of electrostatic fields resulting from the accumulated charge.

A sought-after application of radiation-induced charging is the charge evolution of regolith dust grains on airless bodies, where dielectric grains are simultaneously exposed to solar photons and solar-wind (SW) plasma. Understanding this process is critical both for advancing our knowledge of fundamental planetary surface processes and for addressing practical challenges in space exploration \cite{Katzan1991}. Airless bodies possess virtually no atmosphere and are therefore exposed to solar ultraviolet (UV) radiation, which produces photoelectrons (PEs), as well as to SW plasma. For example, regolith grains on the Moon's surface become electrically charged \cite{Willis1973,Hazra2021, Popel2018}. The behavior of these charged grains influences several key processes, including modification of the local plasma environment, which affects dust transport and adhesion and, in turn, impacts the performance and longevity of equipment such as solar panels, spacesuits, and optical instruments \cite{Goodwin2002, Gaier2007, Colwell2007, Christoffersen2008}. Regolith-grain mobility also has important implications for in situ resource utilization, instrument degradation, and habitat contamination. The charging of regolith grains further affects the performance of electrodynamic dust shields and filtration devices, and electrostatic adhesion to lunar exploration systems \cite{Romero_2022,Romero_2023}. 

The net charge of regolith dust grains is induced by the incident radiation, which includes electrons, ions, generated PEs and other secondary particles. Between neighboring grains, intergrain micro-cavities are formed, allowing for a region where charge can accumulate on opposing surfaces. This localized charge buildup can trigger dust lofting or even dielectric breakdown \cite{Flanagan2006, Sheridan2011, Wang2020, Wang2016, Wang2018}. To quantify these effects, Wang et al.\ \cite{Wang2016} introduced the patched-charge model, which emphasizes the role of micro-cavities in facilitating localized charge accumulation on airless planetary bodies such as the Moon. The patched-charge model considers a portion of a grain within the regolith bed that emits PEs and/or secondary electrons into a cavity, resulting in repulsive forces between surrounding grains \cite{Wang2016}. Complementary to this work, Zimmerman et al.~\cite{Zimmerman_2016} developed a purely analytical model for charge buildup on hexagonally packed spherical grains and showed that electric fields can reach strengths of MV/m in less than one lunar day—sufficient to induce dielectric breakdown \cite{Zimmerman_2016}. However, these analytical models neglect the effects of non-spherical and asymmetrical grain geometries and lack the capability to explore dependencies on grain composition.

In this article, we model charge accumulation within micro-cavities of regolith grains on airless bodies using a KMC framework built in Geant4 (GEometry ANd Tracking) \cite{agostinelli2003geant4,allison2006geant4,allison2016recent}. Here, KMC refers to time-stepped stochastic particle-transport kinetics coupled to a time-evolving electrostatic state, rather than a conventional KMC algorithm, such as residence-time \cite{Fichthorn_1991,Voter2007}, Gillespie \cite{Gillespie1976,Gillespie1977}, or n-fold-way event-list KMC \cite{Bortz1975}. In this framework, Geant4 samples the microscopic scattering and interaction processes, including photoelectric absorption, ionization, elastic scattering, secondary-electron production, implantation, and backscattering, while the resulting deposited charge is accumulated to evolve the macroscopic electrostatic field. This approach enables multiscale simulations with arbitrary geometries and material compositions by connecting microscopic grain-scale electron emission, reabsorption, and SW interactions to the global evolution of charge accumulation. We benchmark our Geant4-based code, called \texttt{g4chargeit}, against a simple stacking of spheres \cite{Zimmerman_2016} and the patched-charge model \cite{Wang2016}. This work provides an open-source, all-in-one software package for the broader scientific community to self-consistently model time-evolving charging phenomena. 

The remainder of the paper is organized as follows. Section~\ref{sec:methods} describes the simulation framework, and Section~\ref{sec:implementation} details the implementation and execution. Section \ref{sec:results} presents grain stacking configurations of increasing complexity and demonstrates the capabilities of the code. The paper concludes with a discussion of limitations and potential applications beyond space science.

\section{Simulation Framework}\label{sec:methods}

Geant4 is an open-source, C++-based simulation toolkit used to model particle scattering and transport through matter, with applications in high-energy, nuclear, medical, and space physics \cite{agostinelli2003geant4,allison2006geant4,allison2016recent}. Our MC framework for time-dependent simulations of electrostatic fields is built in Geant4 (version 11.3.0). The following extensions to Geant4 are incorporated into \texttt{g4chargeit}: (i) Geometry Description Markup Language (GDML) for importing complex geometries and computer-aided design (CAD)-based structures \cite{Chytracek2006}, (ii) Open Multi-Processing (OpenMP) for parallelizing the simulations \cite{openMP}, (iii) ROOT for data storage and post-simulation analysis \cite{root}, (iv) General Particle Source (GPS) for defining isotropic or anisotropic radiation sources, and (v) g4pbc for implementing periodic boundary conditions (PBCs). Building on these extensions, \texttt{g4chargeit} includes a custom class, \texttt{AdaptiveSumRadialFieldMap.cc}, which adaptively computes the electric field from local charges determined in the previous MC simulation iterations.

\subsection{Iterative Approach for Time-Evolved Electric Field} 

The Geant4 toolkit is typically limited by the fact that it does not explicitly model electrodynamics. As a result, particle interactions with the material and electric field are not inherently self-consistent, preventing individual time-steps and trajectories from influencing one another. To circumvent this limitation, we re-initialize the MC simulation at each discretized time step to include the electric-field contributions of charged particles from earlier steps. Specifically, an initial Geant4 simulation is performed for the zeroth iteration ($n = 0$), in which particle transport is modeled based on the probabilistic nature of ionization and scattering processes inside the grains (\ref{app:physics}). Then, for each subsequent iteration, the simulation is re-initialized with the deposited charges of the previous iterations, such that the pre-existing electric field influences subsequent particle trajectories. 

\begin{figure}[!htt]
\centering
\includegraphics[width=\textwidth]{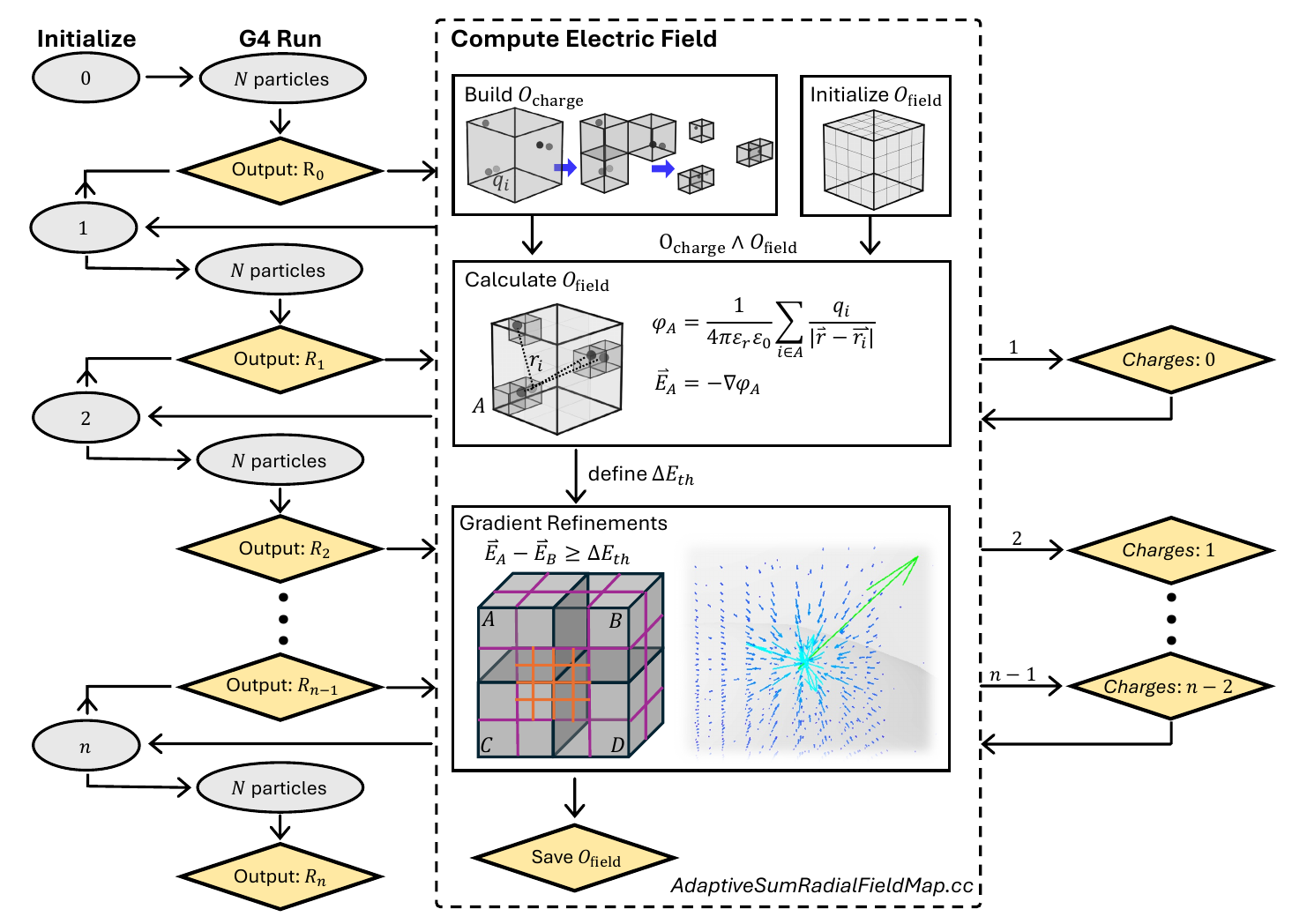}
\caption{Integration of stochastic Monte Carlo simulations provided by Geant4 into the time-evolving, self-consistent simulation framework of \texttt{g4chargeit}. $N$ particles are sampled and transported through the geometry for the $n$-th iteration. The particle trajectories are saved in the ROOT file (denoted as $R_i$), and deposited charges are aggregated into a master charge list (denoted as \textit{Charges}), which is used to compute the electric field. The custom \texttt{AdaptiveSumRadialFieldMap.cc} class (dashed box) is structured as follows: a charge octree, $O_{\mathrm{charge}}$, is constructed, followed by a field octree, $O_{\mathrm{field}}$, generated using a Barness–Hut approximation to efficiently evaluate the electric field (resulting in a scalar potential $\varphi_A$ with a corresponding field $\vec E_A$ for a specific octree voxel $A$). The field octree is adaptively binned if the field gradient between neighboring cells exceeds the threshold $\Delta E_{th}$. An example of the refined $O_{\mathrm{field}}$ is shown as an inset, where the density of the pre-computed electric-field vectors (arrows) scales as the field approaches the deposited charge (center of the image). The kinetic process is repeated for $n$ iterations by self-consistently updating the charge distributions and field maps. }
\label{fig:methods-figure1}
\end{figure}

Figure~\ref{fig:methods-figure1} provides an overview of our kinetic, self-consistent simulation framework. The custom class \texttt{AdaptiveSumRadialFieldMap.cc} calculates the electric field by summing the point-charge contributions from $N$ charges embedded within the grains from prior iterations, using adaptively binned octrees (described in Section~\ref{sec:mesh}). For the $i$-th embedded particle with charge $q_i$ at position $\mathbf{r}_i$, the electric potential contribution is
\begin{equation}\label{eq:potential}
    \varphi_i(\mathbf{r}) = \frac{1}{4\pi \varepsilon_r \varepsilon_0} \frac{q_i}{|\mathbf{r}- \mathbf{r_i}|},
\end{equation} 
where $i \in \mathbb{Z}:[1, N]$, $\varepsilon_r$ is the dielectric constant, and $\varepsilon_0$ is the vacuum permittivity. Here, $\varepsilon_r$ is determined by the material at the field-evaluation point: $\varepsilon_r=1$ in vacuum and the material-specific relative dielectric constant is used within the dielectric grains (e.g., $\varepsilon_r=3.9$ for SiO$_2$ \cite{Volksen_2009}). The treatment of voxels intersecting multiple materials is described in Section~\ref{sec:dielectric-treatement}. At the end of the $n$-th iteration, the locations of all newly deposited charges are recorded and appended to a master list containing the charge-distribution snapshot from iteration $n-1$. With each successive iteration, this master list is loaded into memory, and the total electric field $\mathbf{E}(\mathbf{r})$ is recalculated by summing the contributions of all charges in the geometry, providing an iterative model of charge accumulation and the resulting changes in the local electric field. Charge dissipation is also applied for each iteration based on the conductivity and dielectric constant of the material (\ref{app:charge-dissipation}).

\subsubsection{Adaptive Meshing of Electric Field}\label{sec:mesh}

In electrostatics, the electric potential $\varphi$ in a linear dielectric medium is governed by Poisson's equation, 
\begin{equation}
\nabla \cdot \left(\varepsilon_0\varepsilon_r\nabla\varphi\right)=-\rho,
\end{equation} % Equation (2)
where $\rho$ is the volumetric free-charge density. In charge-free regions, this reduces to Laplace's equation, 
\begin{equation}
\nabla \cdot \left(\varepsilon_0\varepsilon_r\nabla\varphi\right)=0.
\end{equation} % Equation (3)
In our code \texttt{g4chargeit}, these equations are not solved directly on a continuum mesh. Instead, the free charge is represented by the discrete particle charges deposited by Geant4, and the electrostatic potential and field are evaluated from their Coulombic contributions using Eq.~\ref{eq:potential}. Charges deposited near a grain surface contribute to the surface free-charge density $\sigma$ used in the charge-continuity and electric-pressure calculations, whereas charges deposited within the grain represent the discrete analogue of a volumetric charge distribution $\rho$. Because the number of deposited particles can be large, a brute-force recomputation of Eq.~\ref{eq:potential} at every tracking step is computationally expensive. This limitation precludes the use of solving Laplace's equation with boundary potentials or Poisson's equation with charge-density distributions, both of which still require repeated iteration over all deposited charges and do not admit tractable closed-form solutions for the complex geometries considered here.

To reduce the computational cost, we represent the deposited charges as point sources on an adaptive octree, $O_{\mathrm{charge}}$, and the electrostatic field is evaluated on a separate adaptive octree, $O_{\mathrm{field}}$ \cite{MEAGHER1982}. The octree construction is illustrated in Fig.~\ref{fig:methods-figure1} (dashed box). $O_{\mathrm{charge}}$ is constructed by assigning deposited charges to voxels, which are subregions of the tree generated within the geometry. These voxels are then recursively split until either a single integer charge remains or the user-specified minimum voxel size is reached. After $O_{\mathrm{charge}}$ is mapped, $O_{\mathrm{field}}$ is initialized on a uniform coarse grid with a user-defined depth that sets the initial voxel size. $O_{\mathrm{field}}$ is populated using the charges stored in $O_{\mathrm{charge}}$, and distant charge clusters are approximated using a Barnes-Hut-type criterion, while nearby charges are resolved explicitly \cite{gan2014efficient}. The field tree is then refined where the variation in the computed field between neighboring cells exceeds the user threshold $\Delta E_{\mathrm{th}}$, concentrating resolution in regions of large field gradients, such as the micro-cavities between grains. Thus, the refinement criterion is a scalar measure of local field variation, not a full tensor-gradient solver. An example of the adaptively meshed electric field is shown in Fig.~\ref{fig:methods-figure1} (inset, gradient refinements box), where the density of electric field computations is concentrated around the embedded charge. Although the discretized electric field formally contains singularities at the locations of individual embedded charges, the electric potential evaluated away from the individual charges is smooth and differentiable. For a sufficiently small binning size, this direct evaluation naturally captures the relevant multipole moments without explicit treatment.

It is important to note that the two octrees have distinct roles. $O_{\mathrm{charge}}$ is a source tree: it stores the deposited charge locations and recursively groups them until the minimum voxel size is reached. $O_{\mathrm{field}}$ is a target/evaluation tree: it stores the locations at which the electrostatic potential and field are precomputed for subsequent particle tracking. Separating these structures allows the deposited-charge representation to remain tied to particle stopping locations while the field map is refined only where the particle trajectories require additional field resolution. During Geant4 tracking, field lookup is performed from the stored $O_{\mathrm{field}}$ map rather than by direct summation over all deposited charges at each particle step. The final electric-field map is called back into the MC framework, overriding the Geant4-defined field class. Particle trajectories in the resulting electric-field topology are integrated using \texttt{G4DormandPrince745}, a Geant4 field solver that employs a high-order adaptive Runge–Kutta stepper based on the efficient Dormand–Prince (DoPri5) method. The integration accuracy is controlled via the \texttt{DeltaOneStep} parameter set within the \texttt{DetectorConstruction} class; in our simulations, this parameter is set to $\SI{0.1}{\mu m}$, ensuring precise tracking of particle motion \cite{DormandPrince1980}.

\subsubsection{Field Evaluation in Mixed-Material Voxels}\label{sec:dielectric-treatement}

In certain cases, a bin in the field octree may contain $N_v$ materials with different dielectric properties. To account for material variations within a single bin, a weight $w_i$, corresponding to the volume occupation fraction, is assigned to each dielectric constant $\varepsilon_i$. The effective dielectric constant of the bin is then given by
\begin{equation}\label{eq:dielectric-weighting}
\varepsilon_r^{-1}
= \sum_{i=1}^{N_v} w_i\,\varepsilon_i^{-1},
\end{equation} % Equation (4)
allowing a first-order approximation of charge polarization. At an ideal dielectric interface, the electrostatic potential is continuous, while the tangential component of the electric field satisfies
\begin{equation}
    \mathbf{n}\times(\mathbf{E}_2-\mathbf{E}_1)=0,
\end{equation} % Equation (5)
and the normal component of the electric displacement satisfies
\begin{equation}
    \mathbf{n}\cdot(\varepsilon_2\mathbf{E}_2-\varepsilon_1\mathbf{E}_1)=\sigma.
\end{equation} % Equation (6)
In our implementation, these interface conditions are not imposed explicitly as boundary conditions. Instead, material-dependent dielectric constants are incorporated through the field calculation, with $\varepsilon_r=1$ in vacuum and the corresponding material value within each grain. For field voxels that intersect more than one material, Eq.~\ref{eq:dielectric-weighting} provides a volume-weighted effective dielectric constant as a first-order treatment of the dielectric interface. Thus, the adaptive field representation approximates the dielectric response at material boundaries without explicitly solving the continuum jump conditions. Consequently, the effective dielectric treatment avoids numerical discontinuities associated with abrupt voxel-scale changes in material properties while providing a first-order approximation to dielectric response at material interfaces.

\subsubsection{Periodic Boundary Conditions of Particle Trajectories}\label{sec:pbc}

Periodic boundary conditions are applied only in the lateral $x$--$y$ directions for particle trajectories. Particles exiting one lateral boundary are reintroduced through the opposite boundary, representing a periodically repeated grain structure. The deposited charges from these trajectories are then used to update the electrostatic field in the next iteration, so the field is likewise periodic over the lateral simulation cell. The vertical direction is not periodic. Particles leaving the Geant4 simulation cell through the vertical boundaries are no longer tracked. Thus, the present grain-scale simulations do not account for PEs that escape the simulation volume and subsequently return to the surface through interactions with the larger-scale PE sheath.

\subsection{Probabilistic Scattering and Interaction Physics}\label{sec:scattering-examples}

\begin{figure}
\centering
\includegraphics[width=\textwidth]{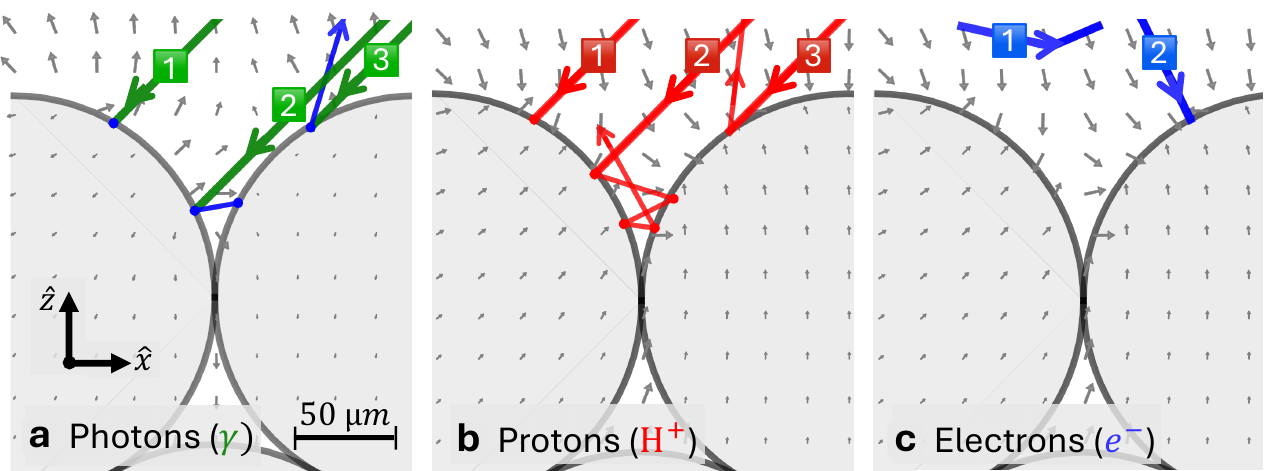}
\caption{Cross-section of hexagonally packed grains (with radius of $\SI{100}{\mu m}$) overlaid with the electric-field map (gray arrows, scaled by field magnitude) in the $xz$-plane. Representative particle trajectories are shown for (a) photons undergoing a PE event, (b) $\SI{1}{keV}$ protons, and (c) low-energy electrons (isotropic in angle). The electric-field maps for photons (a) and SW (b,c) are shown at iterations $n=39$ and $n=17$, respectively, with a dielectric constant of $\varepsilon_r=3.9$. }
\label{fig:methods-figure2}
\end{figure}

Geant4 enables explicit simulation of individual ionization and scattering interactions, providing a versatile toolkit for modeling the production and subsequent propagation of secondary particles. Our approach differs from PIC methods in that the particles are individually tracked, while the electric field is computed separately from the accumulated point charges using an adaptively refined field map (see Section~\ref{sec:mesh}). Here, we demonstrate the particle-tracking capabilities by considering species relevant to the space environment: low-energy solar photons, as well as monoenergetic protons and low-energy electrons from the SW plasma \cite{Willis1973}. For simplicity, we consider hexagonally packed spheres, each with a radius of $\SI{100}{\mu m}$ and composed of \ce{SiO2} (with $\varepsilon_r=3.9$ \cite{Volksen_2009}). Figure~\ref{fig:methods-figure2}a shows the photon case with 39 prior field-update iterations, while Fig.~\ref{fig:methods-figure2}b,c show the SW case with 17 prior field-update iterations. In each iteration, the electric field is recomputed after a batch of $10^6$ photons or $8\times10^4$ SW particles, respectively. Because the SW case involves charged incident particles, the field evolves more rapidly and is updated more frequently than in the photon case.

\paragraph{Photoelectron Trajectories} Approximately 85\% of the photons, sampled from the solar spectrum with an incident angle of 45$^\circ$ (Fig.~\ref{fig:verification-cases}d), undergo the photoelectric effect in our \ce{SiO2} spheres. Of these events, three dominant outcomes are observed with the following probabilities: (1) PE stops near the surface of the sphere (occurring $\sim$97\%), (2) PE scatters out of its originating volume and strikes a neighboring sphere (occurring $\sim$2\%), or (3) PE backscatters and escapes the simulation volume (occurring $\sim$1\%). An example of each scenario is depicted in Fig.~\ref{fig:methods-figure2}a. The photon labeled 1 in Fig.~\ref{fig:methods-figure2}a impacts the grain with an energy of $\SI{233}{eV}$ and penetrates to a depth of $\SI{334}{nm}$ before undergoing the photoelectric effect. The resulting PE, with an energy of $\SI{129}{eV}$, travels $\SI{0.5}{nm}$ before stopping near the surface of the sphere. In this event, an inner-shell electron or Auger electron is emitted with an energy of $\SI{87}{eV}$ and stops within $\SI{0.8}{nm}$ of the initial photoemission site. In contrast, the photon labeled 2 has an initial energy of $\SI{68}{eV}$ and interacts within $\SI{1.5}{nm}$ of the \ce{SiO2} surface, producing a PE with an energy of $\SI{55}{eV}$. This PE traverses $\SI{3.6}{nm}$ of \ce{SiO2} and exits its originating volume with an energy of only $\SI{0.6}{eV}$. Due to the strong electric fields within the cavity between spheres, the PE subsequently loses an additional $\SI{0.4}{eV}$ before entering a neighboring sphere (blue trajectory in Fig.~\ref{fig:methods-figure2}a), where it stops within $\SI{0.4}{nm}$. Finally, a $\SI{41}{eV}$ photon, labeled 3 in Fig.~\ref{fig:methods-figure2}a, undergoes the photoelectric effect within $\SI{0.4}{nm}$ upon interacting with the sphere, producing a $\SI{27}{eV}$ PE that backscatters and escapes the sphere. 

\paragraph{Proton Trajectories} The $\SI{1}{keV}$ protons from the SW plasma, incident at a 45$^\circ$ solar angle, undergo Columbic scattering in \ce{SiO2} and typically implant within $\sim$$\SI{18}{nm}$ (e.g., proton trajectory labeled as 1 in Fig.~\ref{fig:methods-figure2}b). This implantation depth is consistent with results from other simulation codes and experimental results (implantation depth of $\sim$$\SI{22}{}\pm\SI{13}{nm}$ reported in Ref.~\cite{farrell_statistical_2017}). The proton labeled 2 in Fig.~\ref{fig:methods-figure2}b interacts with one sphere, losing $\SI{660}{eV}$ before scattering out and bouncing between the spheres until it escapes the cavity with an energy of $\SI{244}{eV}$ after traveling a total distance of $\SI{16}{nm}$ in both spheres. The third proton shown in Fig.~\ref{fig:methods-figure2}b gains $\SI{1}{eV}$ while interacting with the strong electric fields near the highly illuminated faces of the spheres, then deposits $\SI{950}{eV}$ over $\SI{14}{nm}$ before backscattering out of the sphere and exiting the simulation volume.

\paragraph{Electron Trajectories} Thermal electrons ($<$$\SI{80}{eV}$), incident isotropically in angle, are introduced simultaneously with the $\SI{1}{keV}$ protons to model the SW plasma. The electron labeled 1 in Fig.~\ref{fig:methods-figure2}c illustrates a case in which a $\SI{1.4}{eV}$ electron does not interact with the sphere and is instead deflected by the electric field. The percentage of deflected electrons that do not interact with \ce{SiO2} increases with total fluence because of the stronger electric fields that develop within the geometry (e.g., for $n=17$ in Fig.~\ref{fig:methods-figure2}c, $\sim80\%$ of incident electrons are deflected, compared with 0\% for $n=0$). In contrast, the higher-energy electron labeled 2 in Fig.~\ref{fig:methods-figure2}c, with an energy of $\SI{53}{eV}$, penetrates the sphere and implants at a depth of $\SI{1.2}{nm}$.

\section{Implementation}\label{sec:implementation}

The simulation framework is designed for geometries filled with amorphous dielectric materials (e.g., \ce{SiO2} spheres in vacuum). The charge-accumulation and electric-field calculations assume that the deposited charge has no mobility. Instead, charge leakage from each voxel in $O_{\mathrm{field}}$ is treated macroscopically through the continuity equation for surface charge density (Eq.~\ref{eq:charge_diff} in \ref{app:charge-dissipation}). This approximation is reasonable for dielectric materials with low electrical conductivities (e.g., $\vartheta\sim10^{-13}\ \Omega^{-1}\text{m}^{-1}$). Additionally, the adopted physics list (\ref{app:physics}) is optimized for low-energy photons and electrons, typically below $\sim$$\SI{300}{eV}$, and for ions in the $\SI{}{keV}$-range. Accurate tracking down to the $\SI{}{nm}$-scale (as described in Section~\ref{sec:scattering-examples}) provides a direct connection between microscopic ionization and scattering processes and macroscopic charge accumulation on grain surfaces. 

\subsection{Computational Resources}\label{sec:hpc-resources}

The simulations are performed on a high-performance computing (HPC) cluster. The overall computational load increases with world size (Fig.~\ref{fig:methods-figure1}), since a larger world store more octree voxels in memory; and for each iteration, the construction of the field octree scales as $\mathcal{O}(M \log M)$, where $M$ is the number of deposited charges in the geometry \cite{gan2014efficient}. Each iteration runs on a single node using 24 threads and requires at least 8~GB of memory per process, executed on a central processing unit. SLURM (Simple Linux Utility for Resource Management) is used as the workload scheduler and supports batch scripting. \texttt{g4chargeit} produces an executable in which iteration-specific commands are defined via macros and managed by Python scripts that automate the workflow. Jobs are submitted through SLURM such that each job begins after the previous one completes. The primary bottlenecks arise from inter-dependent threads, which limit parallelization efficiency, and from input/output read-write speeds associated with data handling. \ref{app:HPC-runtimes} shows the simulation run times for the results presented here, carried out on the Georgia Tech Partnership for an Advanced Computing Environment (PACE).

\subsection{Post-Processing of Results}

The output files of our simulation tool include text files containing the electric field at each point of the adaptive octree and ROOT files containing particle trajectories within the geometry (see Appendix A of Ref.~\cite{Vira2023}). In Python, \texttt{pyvista} \cite{Sullivan2019} is used to visualize the field maps, and pyROOT \cite{pyROOT} is used to parse the ROOT files, which record all interactions experienced by each particle within the geometry. For example, the forces between neighboring grains can be calculated using both the field map and the processed ROOT data. The $i$-th component of the electric pressure, used as a measure of the force between grains, is given by
\begin{equation}
f_i = \sigma \, (\mathbf{E} \cdot \mathbf{n})n_i.
\label{eq:fx_component}
\end{equation} 

\section{Verification}\label{sec:results}

We specifically examine the charging of a group of grains within a regolith bed on the Moon for two irradiation scenarios. In one scenario, we simulate solar photons incident at 45$^\circ$. In the other, we simulate the SW plasma, predominantly composed of $\SI{1}{keV}$ protons (also incident at 45$^\circ$) and low-energy electrons (with energies of $\sim\SI{10}{eV}$, isotropic in angle) \cite{Willis1973,farrell_dust_2023}. Our simulations approximate the lunar dayside under quiet solar conditions, where the PE current density is approximately 15 times greater than the SW proton current density \cite{Zimmerman_2016}. We provide quantitative benchmarking of our \texttt{g4chargeit} simulations against Ref.~\cite{Zimmerman_2016} and qualitative comparison with experimental results from Ref.~\cite{Wang2016}. The present verification is focused on grain-scale charging behavior and does not constitute validation of a self-consistent plasma-sheath model. The following subsections describe the geometry and simulation parameters used for verification. The grains in each geometric configuration are filled with \ce{SiO2} ($\varepsilon_r=3.9$ \cite{Olhoeft1975}, except $\varepsilon_r=1$ for comparison with Ref.~\cite{Zimmerman_2016}) and placed in a galactic environment to model charging dynamics under realistic planetary conditions. Our MC approach inherits the validated cross-sections implemented in Geant4 \cite{Li2022, Albqoor2023}, making it highly versatile and enabling straightforward modification of material composition and grain geometry without altering the underlying physics models.

\subsection{Geometry for Benchmarking}

\begin{figure}[!b]
\centering
\includegraphics[width=0.8\textwidth]{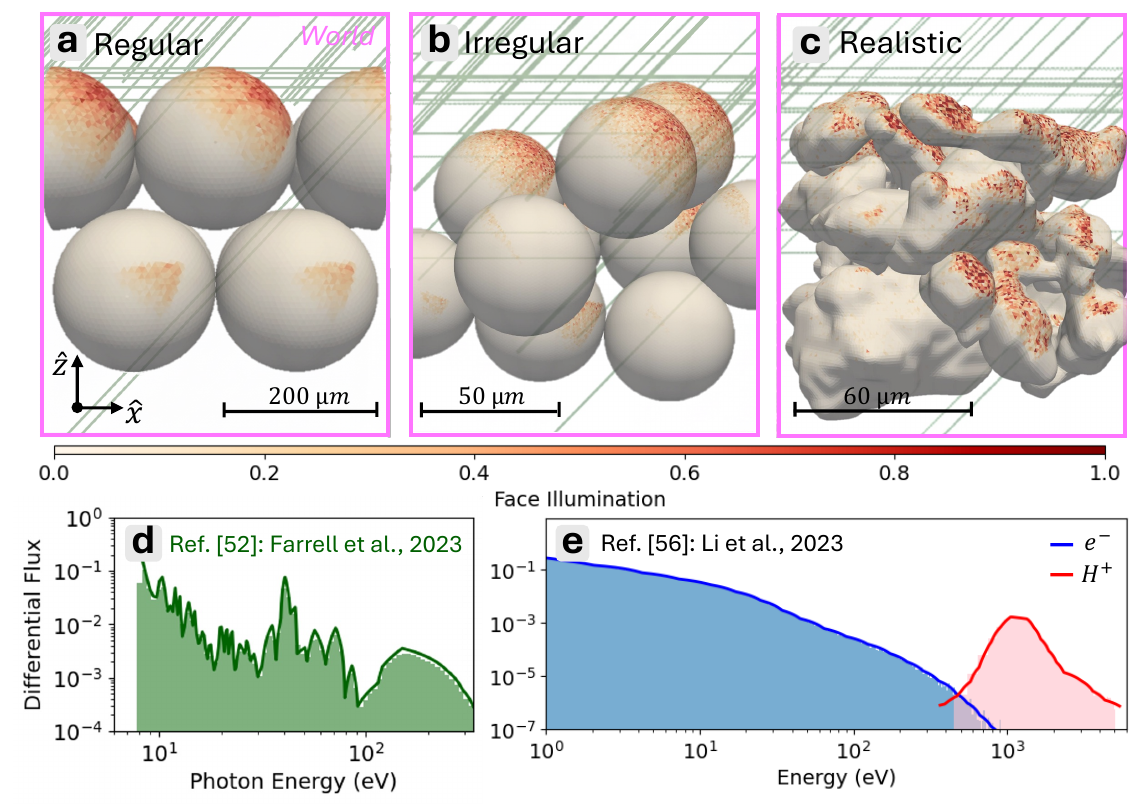}
\caption{Face illumination of (a) regularly packed grains, (b) irregularly packed grains, and (c) a realistic grain configuration. Particle trajectories of photons incident at 45$^\circ$ are shown in green. The horizontal green lines are due to PBC in the $x$-direction, where particles re-enter the box after exiting through the sides. (d) Photon energies are sampled from the differential flux at lunar dayside under quiet solar conditions (Ref.~\cite{farrell_dust_2023}). (e) Particles from the SW plasma are sampled from the differential flux of electrons (blue) and protons (red) (Ref.~\cite{liFormationLunarSurface2023}). SW electrons are isotropic with an average energy of $\sim$$\SI{12.7}{eV}$, and SW protons are incident at 45$^\circ$ angle with an average energy of $\sim$$\SI{1.2}{keV}$.}
\label{fig:verification-cases}
\end{figure}

We consider the following input geometry configurations:

\paragraph{Regularly Packed Grains} For the simplest configuration, we consider a hexagonally packed arrangement of spheres in the (010)-plane (cross-section in the $xz$-plane shown in Fig.~\ref{fig:methods-figure2}). Each sphere has a radius of $\SI{100}{\mu m}$ and is composed of \ce{SiO2} with a density of $\SI{2.2}{g/cm^3}$. We assume an amorphous and homogeneous \ce{SiO2} composition for simplicity; however, this can be readily modified in \texttt{g4chargeit} within the \texttt{DetectorConstruction} class using standard Geant4 commands \cite{agostinelli2003geant4}. The simulation volume is reduced to the smallest periodic unit surrounding the central sphere (Fig.~\ref{fig:verification-cases}a, pink box), and PBCs are enabled in the $xy$-plane to simulate an unbounded lateral repetition of grains. The stacking configuration and sphere radius are chosen to replicate the structure in Ref.~\cite{Zimmerman_2016}, allowing for a quantitative comparison with our results.

\paragraph{Irregularly Packed Grains} We increase the geometric complexity by considering irregularly packed spherical grains (Fig.~\ref{fig:verification-cases}b) as an intermediate configuration between the regularly packed spheres and the realistic grain arrangement. Each sphere has a radius of $\SI{25}{\mu m}$, chosen to better represent the typical sizes of grains in lunar soil \cite{carrierLunarSoilGrain1973,goguen_2024}. We continue to assume a simple \ce{SiO2} composition but use a density of $\SI{1.9}{g/cm^3}$, reflecting measured densities of lunar simulants \cite{Wang2016}.

\paragraph{Realistic Grain Configuration} The grains are arranged in a realistic grain‐stack configuration (Fig.~\ref{fig:verification-cases}c) that reproduces a geometry derived from synchrotron X-ray microtomography (Beamline 8.3.2, Advanced Light Source) of a planetary simulant. The isolated grains are arranged to fill a $107 \times 121 \times \SI{100}{\mu m}$ volumetric region to resemble regolith; the resulting porosity is $\sim$52\%, in agreement with measured porosity values of lunar soil \cite{carrierLunarSoilGrain1973,goguen_2024}. This configuration also assumes a \ce{SiO2} composition with a density of $\SI{1.9}{g/cm^3}$.

\subsection{Incident Particle Distributions}\label{sec:particle_flux}

We consider two distinct irradiation cases: (1) photon irradiation and (2) SW irradiation. Photons are sampled directly from the solar spectrum (Fig.~\ref{fig:verification-cases}d, Ref.~\cite{farrell_dust_2023}), which is based on the Flare Irradiance Spectral Model (FISM) \cite{Chamberlin2008} using measurements from the Solar EUV Experiment (SEE) \cite{woods2005} and Solar Stellar Irradiance Comparison Experiment (SOLSTICE) \cite{Rottman1993}. The sampled photon energies span $\sim$$\SI{8}{eV}$ to $\SI{350}{eV}$ and uniformly strike the geometry at a $45^\circ$ angle to simulate the average dayside conditions. The photoelectric effect is modeled stochastically (e.g., Fig.~\ref{fig:methods-figure2}a). For the SW case, protons with energies of $\sim$$\SI{1}{keV}$ (red histogram in Fig.~\ref{fig:verification-cases}e, Ref.~\cite{liFormationLunarSurface2023}) are incident at a $45^\circ$ angle, while low-energy electrons are incident isotropically. The SW electrons are sampled from the distribution shown in Fig.~\ref{fig:verification-cases}e (blue histogram, Ref.~\cite{liFormationLunarSurface2023}), with an average energy of $\sim$$\SI{12.7}{eV}$ and a current density five times that of the protons. 

The simulation parameters for each case and configuration are summarized in Table~\ref{tab:simulation-parameters}. To relate the simulated particle fluence to a physically meaningful timescale, we define a lunar equivalent time ($t_{\text{M}}$) for each iteration using the corresponding average dayside current densities on the Moon from Refs.~\cite{Willis1973,Zimmerman_2016}. For photon irradiation, $t_{\text{M}}$ is determined from the number of PEs emitted from the grain surfaces, since photoemission is the charging process of interest, using an average PE current density of $4\times10^{-6}\SI{}{A/m^2}$. For SW irradiation, $t_{\text{M}}$ is instead determined from the number of incident protons, using an average proton current density of $3\times10^{-7}\SI{}{A/m^2}$. The SW electron population is introduced simultaneously with the protons at a current density of $1.5\times10^{-6}\SI{}{A/m^2}$, five times the proton current density, such that the relative electron and proton fluxes are preserved while a single SW timescale is defined by the incident proton flux. Thus, the ``particles at $n=0$'' in Table~\ref{tab:simulation-parameters} correspond to emitted PEs for the photon case and incident protons for the SW case and are used to convert the timestep of each simulation into $t_{\text{M}}$. 

\begin{table}[!ht]
    \centering
    \caption{Lunar equivalent time ($t_{\text{M}}$) for each configuration (regularly packed, irregularly packed, and realistic packing) is calculated using average dayside current densities of $4\times10^{-6}\SI{}{A/m^2}$ for emitted PEs and $3\times10^{-7}\SI{}{A/m^2}$ for SW protons \cite{Zimmerman_2016}, corresponding to a flux of $2.49\times10^{13}\SI{}{1/m^2/s}$ and $0.18\times10^{13}\SI{}{1/m^2/s}$ respectively. Specifically, $t_{\text{M}}$ is obtained by normalizing the number of particles incident on the simulation volume by the area of the illumination plane and the corresponding average dayside flux. }
    \hspace{0.1cm}
    \label{tab:simulation-parameters}
    \begin{tabular}{p{3.4cm}||c|c||c|c||c|c}
        \hline
        & \multicolumn{2}{c||}{Regularly Packed} 
        & \multicolumn{2}{c||}{Irregularly Packed} 
        & \multicolumn{2}{c}{Realistic} \\
        \hline
        Illumination Plane ($\SI{}{\mu m}$)
            & \multicolumn{2}{c||}{400 $\times$ 300} 
            & \multicolumn{2}{c||}{120 $\times$ 105} 
            & \multicolumn{2}{c}{107 $\times$ 121} \\
        \hline\hline
        & $\gamma$ & SW
        & $\gamma$ & SW 
        & $\gamma$ & SW \\
        \hline
        Incident particles per $n$ 
            & {1{,}000{,}000} & {80{,}000}
            & {500{,}000} & {10{,}000} 
            & {300{,}000} & {10{,}000} \\
        \hline
        Fluence per $n$ ($\SI{}{\mu m^{-2}}$)
            & 8.33 & 0.67
            & 39.68 & 0.79
            & 23.21 & 0.77 \\
        \hline\hline
        & PE & \ce{H^+}
        & PE & \ce{H^+} 
        & PE & \ce{H^+} \\
        \hline
        Particles at $n=0$ 
            & 8,446 & 13,060 
            & 3,270  & 1,235 
            & 2,917  & 1,409 \\
        \hline
        \shortstack[l]{$t_{\text{M}}$ per $n$ ($\SI{}{ms}$)}
            & 2.82 & 58.12 
            & 10.40 & 52.35 
            & 9.04 & 58.23\\
        \hline
    \end{tabular}
\end{table}

\subsection{Simulation Results}

\subsubsection{Regularly Packed Grains}

\begin{figure}[!t]
\centering
\includegraphics[width=0.8\textwidth]{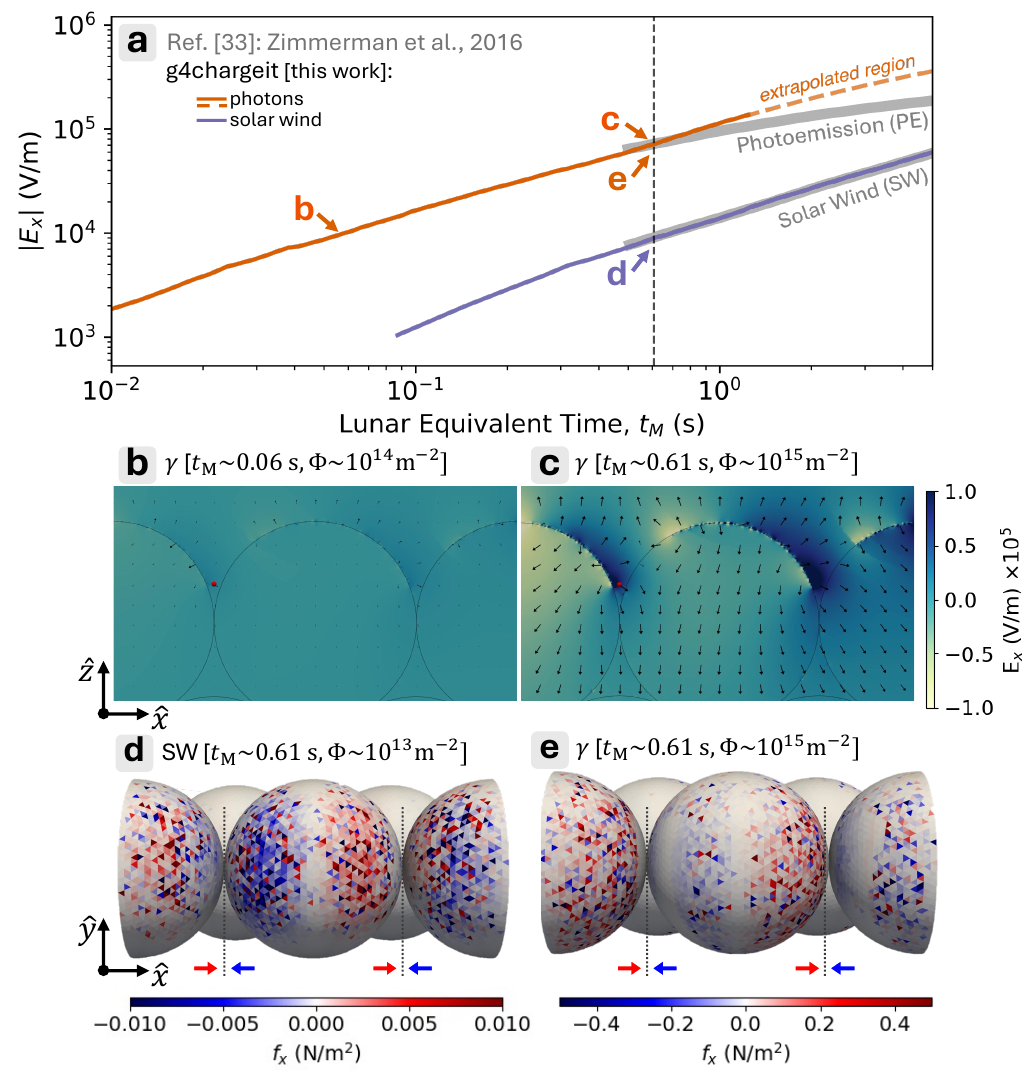}
\caption{(a) $|E_x|$ plotted against the equivalent lunar time $t_{\text{M}}$ for the photon (orange) and SW (purple) cases at the point $\SI{37}{\mu m}$ above the midpoint between the sphere centers; red point in (b) and (c). The photon results are extrapolated using the expected extension (dashed orange line, least-squares fit to Eq.~\ref{eq:charge_diff}). Results from Ref. \cite{Zimmerman_2016} are overlaid for comparison (gray lines). (b,c) Interpolated electric-field vectors in the $xz$-plane at (b) $t_{\text{M}}\sim\SI{0.06}{s}$ and (c) $t_{\text{M}}\sim\SI{0.61}{s}$. The background color shows the $x$-component of the electric field $E_x$. (d,e) Electric pressure ($x$-component) at $t_{\text{M}}\sim\SI{0.61}{s}$ for (d) SW irradiation (equivalent fluence of $\Phi\sim10^{13}\SI{}{m^{-2}}$) and (e) photons irradiation (equivalent fluence of $\Phi\sim10^{15}\SI{}{m^{-2}}$). An attractive force between the spheres is observed for both SW and photon irradiation (adjacent regions with opposite colors, i.e., red–blue).}
\label{fig:benchmark}
\end{figure}

Figure~\ref{fig:benchmark} benchmarks our simulation against the analytical approach presented in Ref.~\cite{Zimmerman_2016}. Figure~\ref{fig:benchmark}a shows the $x$-component of the electric field $|E_x|$ for each iteration at a point $\SI{37}{\mu m}$ above the midpoint between the sphere centers (red point in Fig.~\ref{fig:benchmark}b,c). The simulations are initialized at early times and iteratively advanced, allowing the electric field to evolve gradually until convergence with the analytical model (gray lines in Fig.~\ref{fig:benchmark}a). We find excellent agreement for both the photon and SW cases, which include modeling of the passive current density and charge dissipation (Eq.~\ref{eq:charge_diff}).  For the SW case, the simulations agree within $\sim$4.85\%, indicating that \texttt{g4chargeit} accurately captures charge accumulation in this simple geometry and reliably reproduces the temporal evolution of the electric field under the same irradiation conditions in Ref.~\cite{Zimmerman_2016}. 

However, for photon irradiation, our simulation results cannot be directly compared with those of Ref.~\cite{Zimmerman_2016}. The photoelectric effect is not explicitly calculated in Ref.~\cite{Zimmerman_2016}; instead, PEs with an average energy of $\SI{1}{eV}$ are isotropically emitted from the illuminated faces. This assumption produces a population of low-energy electrons whose trajectories are readily perturbed by the strong electric fields that develop between the grains, ultimately contributing to the nonlinear growth and eventual saturation of the electric field (top gray line in Fig.~\ref{fig:benchmark}a). In our simulations, photons are sampled directly from the solar spectrum (Fig.~\ref{fig:verification-cases}d), and the photoelectric effect is modeled explicitly (e.g., Fig.~\ref{fig:methods-figure2}a), removing the need to impose assumptions about the secondary yield and instead utilizing the validated cross-sections implemented in Geant4 \cite{Li2022, Albqoor2023}. Because the energy of PEs is $E_{\gamma} - \phi$, where $\phi$ is the work function ($\sim$$\SI{8}{eV}$), the generated PEs can have energies as high as $\sim$$\SI{340}{eV}$, significantly larger than the $\SI{1}{eV}$ PE population in Ref.~\cite{Zimmerman_2016}. These more energetic PEs require stronger electric fields to deflect them, which, in turn, leads to an increased $E_x$ at saturation in our results (extrapolated region, orange dashed line in Fig.~\ref{fig:benchmark}a). % as these energetic electrons can still escape through the potentials established on adjacent grains. 
As a result, the electric field is expected to plateau at $8.4\times10^5\SI{}{V/m}$ at $t_\text{M}=\SI{120}{s}$, when the charge dissipation term starts to dominate the surface charge variation (Eq.~\ref{eq:charge_diff}). This value is $\sim$$2\times$ larger than that reported in Ref.~\cite{Zimmerman_2016} of $|E_x|=3.6\times10^5\SI{}{V/m}$ at $t_\text{M}=\SI{120}{s}$, but remains within the expected range for a constant conductivity of $\vartheta\sim10^{-13}\ \Omega^{-1}\text{m}^{-1}$. Nevertheless, our simulation results are in close agreement with Ref.~\cite{Zimmerman_2016} (Fig.~\ref{fig:benchmark}a).

%\ref{app:PEs} provides the distributions and statistics of the generated photoelectrons for regularly-packed spheres.

The attractive or repulsive force between neighboring spherical grains is quantified using the electric pressure (Eq.~\ref{eq:fx_component}). The $x$-component of the electric pressure is shown in Fig.~\ref{fig:benchmark}d,e, where adjacent regions of opposing sign (red–blue) correspond to an attractive interaction. For the regularly packed spherical grains, we find that the grains experience a net attractive force, consistent with Ref.~\cite{Zimmerman_2016}. However, both our simulations and the grain-scale model in Ref.~\cite{Zimmerman_2016} stand in conflict with experimental observations, which instead report a repulsive force inside micro-cavities that may contribute to grain lofting \cite{Wang2018,hood_laboratory_2018,Wang2020}. This discrepancy suggests that regularly packed spheres may not capture key physical effects responsible for the experimentally observed repulsion. The patched-charge model has previously explored the importance of micro-cavities using both experimental and numerical approaches \cite{Wang2016,Wang2018,Wang2020}. Although it represents an improvement over the conventional uniform shared-charge assumption, it still relies on simplifications—most notably idealized grain shapes and the neglect of transient charge fluctuations and stochastic electron trajectories—that may limit its applicability under more realistic conditions. To investigate whether geometric complexity, such as micro-cavities, can produce the repulsive regions observed in experiments, we apply our self-consistent MC framework to simulate charge accumulation in irregularly packed grains.

\subsubsection{Irregularly Packed Grains}

\begin{figure}[!ht]
\centering
\includegraphics[width=0.9\textwidth]{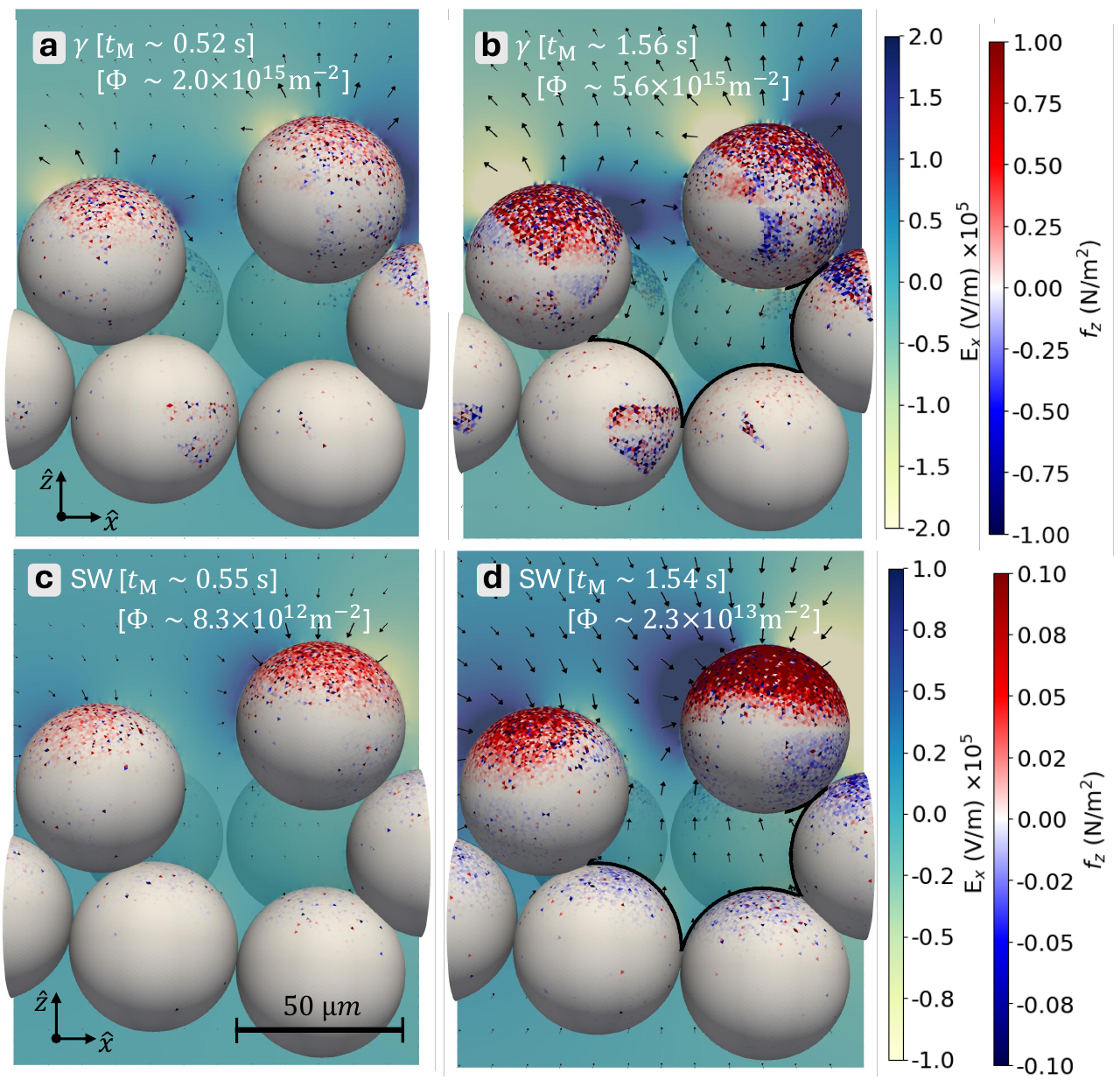}
\caption{Simulation results for irregularly packed grains under photon irradiation at (a) $t_{\text{M}}\sim\SI{0.52}{s}$ (corresponding fluence $\Phi\sim2.0\times10^{15}\SI{}{m^{-2}}$) and (b) $t_{\text{M}}\sim\SI{1.56}{s}$ ($\Phi\sim5.6\times10^{15}\SI{}{m^{-2}}$), followed by the SW case of (c) $t_{\text{M}}\sim\SI{0.55}{s}$ ($\Phi\sim8.3\times10^{12}\SI{}{m^{-2}}$) and (d) $t_{\text{M}}\sim\SI{1.54}{s}$ ($\Phi\sim2.3\times10^{13}\SI{}{m^{-2}}$). Each panel shows the $x$-component of the electric field ($E_x$, yellow–blue colormap), the $z$-component of the electric pressure ($f_z$, seismic colormap), and the electric-field vector (black arrows) in the $xz$-plane.} 
\label{fig:irregular-grains}
\end{figure}

Figure~\ref{fig:irregular-grains} illustrates charge accumulation on irregularly packed spherical grains for photon (Fig.~\ref{fig:irregular-grains}a,b) and SW (Fig.~\ref{fig:irregular-grains}c,d) irradiation. The $z$-component of the electric pressure $f_z$ is shown on the surface of each sphere, while the background color represents the $x$-component of the electric field $E_x$, and the vectors indicate the electric field in the $xz$-plane (scaled by the field magnitude). These results address the question posed earlier: can repulsive forces emerge within micro-cavities for different grain packings? In contrast to the attractive forces observed for regularly packed grains (Fig.~\ref{fig:benchmark}d,e), the irregular configuration exhibits distinct repulsive regions (i.e., red–red or blue–blue in Fig.~\ref{fig:irregular-grains}b,d). As the system evolves, charged particles are scattered into micro-cavities (black-outlined regions in Fig.~\ref{fig:irregular-grains}b,d), generating electrostatic repulsion between neighboring grains. This repulsion is particularly evident for the SW case (Fig.~\ref{fig:irregular-grains}d), where substantial charge buildup within the cavity produces forces that can eventually push the spheres apart. At an equivalent time for photon irradiation (Fig.~\ref{fig:irregular-grains}b), the repulsive force is weaker due to the lower electric-field strength within the cavity. We find that the irregular packing can induce repulsive forces between spherical grains, whereas regular hexagonal packing (Fig.~\ref{fig:benchmark}d,e) produces symmetric pressure patterns that induce attraction. The repulsive force observed within the micro-cavity in Fig.~\ref{fig:irregular-grains} qualitatively agrees with Ref.~\cite{Wang2016}, in which the authors found that exposure to charged particles generates electrostatic repulsion, while exposure to photons alone does not produce comparable field strengths within micro-cavities.

It is important to consider the timescales over which our simulations are valid, since our geometry remains static during the time-discretized MC runs. Consequently, this framework cannot be used directly to simulate grain lofting; instead, it captures the precursor electric fields that develop within micro-cavities before the forces become large enough to move the spheres. In Fig.~\ref{fig:irregular-grains}d, $f_z$ reaches a maximum of $\sim$$\SI{0.1}{N/m^2}$ within a micro-cavity, corresponding to an order-of-magnitude force of $F_z \sim10^{-13}\SI{}{N}$. For comparison, the lunar gravitational force on a silica sphere of $\SI{50}{\mu m}$ diameter is $F_g = 2.0 \times 10^{-10}\SI{}{N}$. Thus, $F_z$ would need to increase by roughly three orders of magnitude before lofting could occur. Therefore, on the timescales of seconds, $F_z \ll F_g$, and the simulations shown in Fig.~\ref{fig:irregular-grains} remain well within the regime in which the assumption of static geometry is valid. 

\subsubsection{Realistic Grain Configuration}

\begin{figure}[!ht]
\centering
\includegraphics[width=0.9\textwidth]{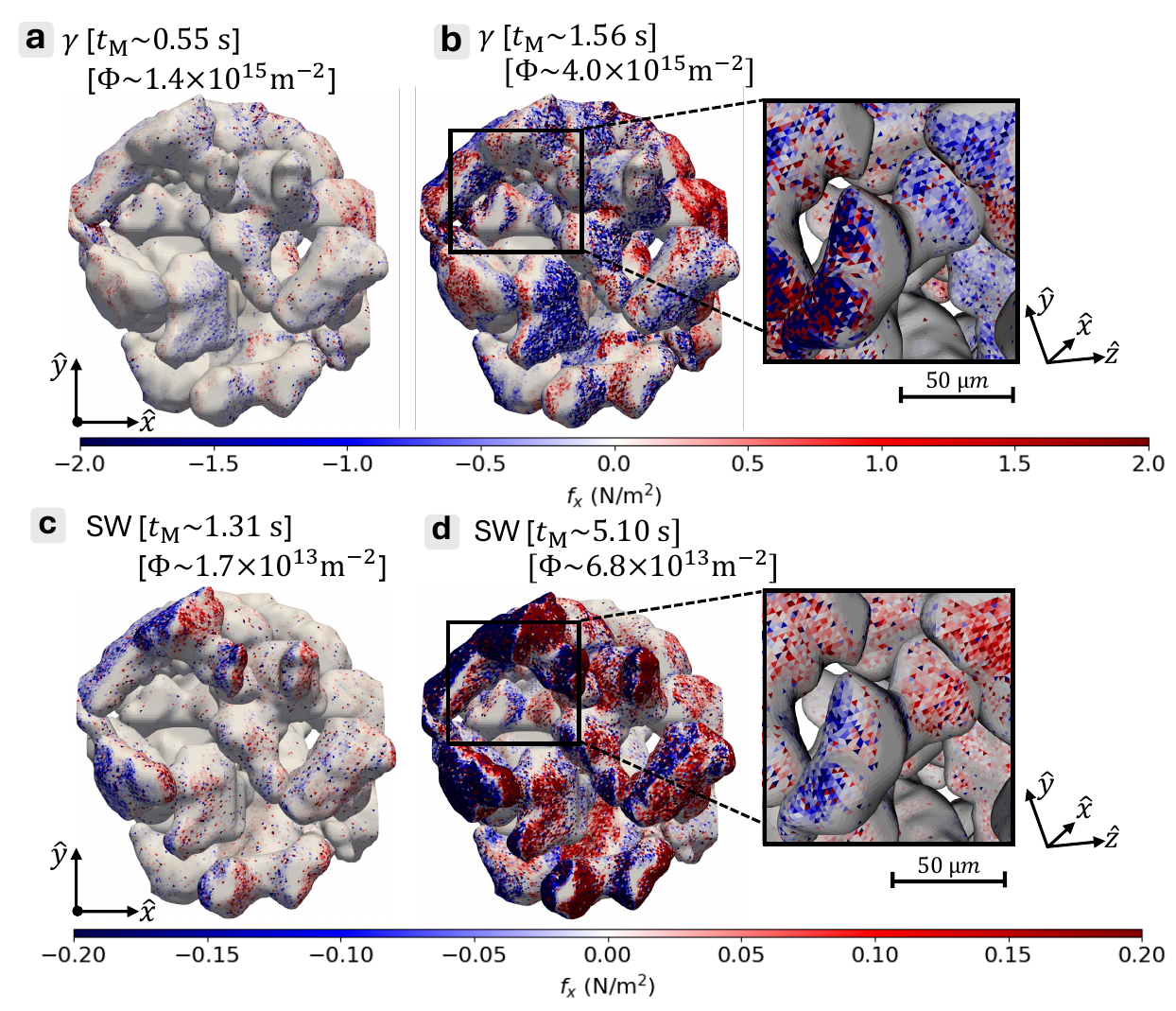}
\caption{Electric pressure ($x$-component) for the realistic packing of grains under photon irradiation at (a) $t_{\text{M}}\sim\SI{0.55}{s}$ (corresponding fluence $\Phi \sim 1.4\times 10^{15} \SI{}{m^{-2}}$) and (b) $t_{\text{M}}\sim\SI{1.56}{s}$ ($\Phi \sim 4.0\times 10^{15} \SI{}{m^{-2}}$), with the inset showing the electric pressure within a micro-cavity. The analogous results for SW irradiation are shown at (c) $t_{\text{M}}\sim\SI{1.31}{s}$ ($\Phi \sim 1.7\times 10^{13} \SI{}{m^{-2}}$) and (d) $t_{\text{M}}\sim\SI{5.10}{s}$ ($\Phi \sim 6.8\times 10^{13} \SI{}{m^{-2}}$). The zoomed-in insets in (b) and (d) show the generation of a repulsive force (adjacent regions of the same color, i.e., blue–blue or red–red) for both irradiation scenarios.} 
\label{fig:realistic-grains}
\end{figure}

Figure~\ref{fig:realistic-grains} shows the time evolution of the $x$-component of the electric pressure $f_x$ for the realistic grain configuration, in which charge accumulation within micro-cavity is observed by comparing Fig.~\ref{fig:realistic-grains}a with Fig.~\ref{fig:realistic-grains}b for photon irradiation and Fig.~\ref{fig:realistic-grains}c with Fig.~\ref{fig:realistic-grains}d for SW irradiation. Figure~\ref{fig:realistic-grains}b,d present snapshots of the surface charge distribution at later times, including zoomed-in insets of representative micro-cavities at $t_\text{M}\sim\SI{1.56}{s}$ and $\SI{5.10}{s}$ for photon and SW irradiation, respectively. Interpreting the electrostatic force is challenging for such a complex geometry, as it depends on the local face normal (Eq.~\ref{eq:fx_component}); however, a qualitative understanding of the forces shown in the insets of Fig.~\ref{fig:realistic-grains}b,d can be obtained by considering $f_x$ in the context of the coordinate axes. Because adjacent regions in the insets of Fig.~\ref{fig:realistic-grains}b,d have the same sign (i.e., blue–blue or red–red), neighboring grains experience repulsive interactions, which could increase the porosity of the grain stacking if the system were evolved to longer times. Nevertheless, the observed charge accumulation within micro-cavities is consistent with the patched-charge model \cite{Wang2018}, in which the emission and reabsorption of charged particles generate large surface charge densities and resulting repulsive forces within grain micro-cavities.

Charge accumulation in realistic geometries produces a markedly more complex distribution than in spherical grains. Sharp morphological features introduce micro-cavities that support localized regions of strong repulsion, which may promote dielectric breakdown and increase the likelihood of grain lofting as electric fields concentrate along protrusions and recessed regions. Because the charge distribution becomes spatially asymmetric with irregular grains, the local electric-pressure contributions are no longer canceled by a matching opposing pattern as they are in the regularly packed spheres. As a result, the micro-cavity regions in Fig.~\ref{fig:realistic-grains}b,d exhibit localized repulsion that is dictated by the grain geometry. It leads to a highly nonuniform electric-pressure patterns that fail to form the opposing pressure-pair structures observed in regularly packed spherical grains, resulting instead in localized repulsion dictated by grain geometry. Because our simulation framework is built on Geant4, the open-source code \texttt{g4chargeit} can be readily adapted to study other charging phenomena within lunar science and the broader planetary science community. 

\section{Other Applications}\label{sec:other-app}

The core implementation of \texttt{g4chargeit} functions as a general-purpose KMC tool for studying the charging of dielectric materials. It is readily adaptable to airless planetary bodies through modifications of the grain geometry, material composition, and incident particle distributions. 

\paragraph{Advantage of the Geant4 Monte Carlo Approach} 

The \texttt{g4chargeit} framework combines Geant4 particle transport with a custom electric-field calculation to model electrostatic charging of dielectric materials in complex planetary environments. Geant4 stochastically models the microscopic interactions and trajectories of incident and secondary particles, while the deposited charges are used by our custom field class to calculate the evolving electric field, which is subsequently coupled back into Geant4 to influence particle trajectories. The accumulation of discrete charged particles naturally resolves local charge inhomogeneities, enabling the calculation of spatially varying electric pressure by connecting the microscopic, atomic-scale interactions to the micron-scale charge inhomogeneities. The use of discrete charge deposition and adaptive field resolution is particularly useful for resolving localized charging and short-range interactions within micro-cavities and irregular grain geometries. In addition, the framework supports arbitrary geometries and materials while leveraging the low-energy particle-interaction models and cross-section databases implemented in Geant4 to model complex radiation transport processes, including secondary electron emission and backscattering. The present model focuses on PE emission and charging at the individual-grain scale and does not explicitly model the larger-scale PE sheath that forms above the lunar surface. Consequently, feedback from the sheath, including returning PEs, is not included. Future work involves coupling the grain-scale model presented here to a larger-scale PIC plasma code (e.g., Ref.~\cite{Farrell_2013}) by using its results as input, while also improving the efficiency of our field octree calculation to support a fully integrated multiscale model. The strength of \texttt{g4chargeit} lies in bridging atomistic interaction physics and grain-scale charge evolution while incorporating the complexities of real materials. 

\paragraph{Applications within Planetary Science}
Within the planetary science community, \texttt{g4chargeit} can be used to quantify how the geometry of realistic grains influences charge accumulation and to assess the role of these parameters in the eventual electrostatic lofting of grains. For example, soil porosity, grain size, and material composition can be systematically modified to study their effects on charge accumulation (e.g., comparison with Ref.~\cite{hood_laboratory_2018}). Beyond grain structure and composition, the properties of the incident particle population can also be adjusted to investigate additional charging processes relevant to the Moon and other airless bodies. For instance, solar energetic particles can be modeled; because these particles penetrate deeper into the material and undergo more complex scattering, \texttt{g4chargeit} leverages the Geant4 architecture to stochastically simulate the associated ionization and scattering processes. In such cases, deeper energy deposition makes dielectric screening increasingly important. These examples illustrate the connection that our code establishes between microscopic scattering interactions and macroscopic charge accumulation and electric-field evolution (e.g., Ref.~\cite{Halekas_2009}).

\paragraph{Potential Utility beyond Space Science}
Additional applications include simulating dust mitigation technologies such as electrostatic dust shields, placing limits on dielectric breakdown in materials, modeling spacecraft charging in the presence of external electric or gravitational fields, and extending Geant4’s capabilities for condensed-matter and semiconductor device simulations \cite{g4cmp}. More broadly, the framework is applicable to diverse domains, including photovoltaic devices and electrostatic processes in biological systems such as DNA. 

\section{Concluding Remarks}\label{sec:discussion}
%- Spatially varying Electric Pressure calculated all at once
%- Statics of higher order multipole moments are accounted for
%- Arbitrary geometries can be used without additional computational overhead
%- All physics contributions taken care of by backend
%- Applications outside of E and M.

In this work, we present a KMC approach to model the time evolution of electrostatic charging on dielectric materials. We apply our code, \texttt{g4chargeit}, to simulate the charging of non-spherical dust grains within a lunar regolith bed, a harsh electrostatic environment in which charged dust can significantly influence local plasma conditions, dust transport, and adhesion. By leveraging a MC approach, inherited from Geant4, we model the generation of secondary electrons and the associated electric fields on non-spherical dielectric grains, developing a generalizable and adaptable framework. We show that charge accumulation leads to localized enhancements in surface charge density, resulting in electric-field heterogeneities that strongly influence subsequent particle interactions, secondary emission, and ultimately grain-scale dynamics. By explicitly resolving secondary electron production and time-dependent charge accumulation, the model naturally captures microscopic charging processes that connect to macroscopic behavior. We demonstrate that this multiscale simulation approach can handle complex, heterogeneous regolith structures and provide a versatile code with applications beyond planetary science. 

\section*{Credit authorship contribution statement}
\textbf{Kush P. Gandhi}: Preparation of original draft, Conceptualization, Methodology, Validation. \textbf{Advik D. Vira}: Preparation of revised draft, Formal Analysis, Supervision. \textbf{William M. Farrell}: Manuscript review, Conceptualization, Validation. \textbf{Nikolai Simonov}: Methodology. \textbf{Alvaro Romero-Calvo}: Manuscript review, Methodology, Validation. \textbf{Thomas M. Orlando}: Manuscript review, Funding acquisition, Project administration. \textbf{Phillip N. First}: Manuscript review, Methodology, Funding acquisition. \textbf{Zhigang Jiang}: Manuscript review, Conceptualization, Supervision, Funding acquisition.

\section*{Declaration of Competing Interest}
The authors declare that they have no known competing financial interests or personal relationships that could have appeared to influence the work reported in this paper.

\section*{Acknowledgments}
The authors thank Karl Smith for inspiring the development of our simulation framework, Makoto Asai for numerous insightful discussions on Geant4, and Romain Fonteyne for his invaluable assistance during the early stages of this work. This research was supported by the NASA Solar System Exploration Research Virtual Institute (SSERVI), under cooperative agreement number NNH22ZDA020C (CLEVER, Grant number: 80NSSC23M022). It was also supported in part through research cyber infrastructure resources and services provided by the Partnership for an Advanced Computing Environment (PACE) at the Georgia Institute of Technology, Atlanta, Georgia, USA.

\appendix

\section{Modification to Geant4 Factory Physics Lists}\label{app:physics}

Geant4 has been extended beyond its original high-energy physics scope to describe interactions down to the eV scale \cite{Arce2025} and, in some applications, low-energy processes at molecular scale \cite{Sakata2020}. These developments have expanded the use of Geant4 for low-energy particle interactions, including the photoelectric effect and Compton scattering, in Geant4 version 11.3 \cite{Li2022}. For our simulations, we modify the factory physics list \texttt{FTFP\_BERT\_EMX} by replacing the standard electrostatic (EM) physics package with \texttt{G4EmStandardPhysics\_option4}.
This option employs the low-energy Livermore models to compute interaction cross-sections using evaluated databases, including EPDL97 (Evaluated Photon Data Library) \cite{osti_295438}, EPICS2017 Electron Photon Interaction Cross Section \cite{Cullen2018EPICS2017, Li2022EPICS2017}, EEDL (Evaluated Electron Data Library) \cite{Perkins_1991}, EADL (Evaluated Atomic Data Library) \cite{osti_10121422}, and electron binding energies derived from Scofield’s data \cite{crasemann_6_1975}. These databases, combined with experimental measurements and theoretical models, allow Geant4 to calculate total and sub-shell–resolved photoelectric cross-sections, Compton scattering cross-sections, secondary-particle energy spectra, electron binding energies, and transition probabilities for fluorescence and Auger emission. 

To accurately model low-energy particles, we further configure the EM processes via \texttt{G4EmParameters}. Atomic de-excitation processes—including fluorescence, Auger electron production and cascades, and Particle-Induced X-ray Emission (PIXE)—are enabled, along with the Continuous Slowing Down Approach (CSDA) to model continuous energy loss. The default production thresholds are reduced to permit energy deposition down to $\SI{0}{eV}$ for electrons and $\SI{10}{eV}$ for hadrons. The default range cuts are also adjusted for hadronic interactions to reduce the step-size for the multiple scattering and continuous energy-loss processes using a range factor of $0.02$ for multiple scattering and a range fraction of $0.05$ with a maximum step length of $\SI{10}{nm}$ for energy loss. We validate the stopping range of ions and electrons from Geant4 against Stopping and Range of Ions in Matter (SRIM) \cite{srim_textbook,ziegler2013stopping} and CASINO \cite{casino}, respectively. Together with the low-energy Livermore model and adjusted range cuts for hadronic scattering, these physics choices enable accurate modeling of the low-energy charge-generation and transport processes associated with SW irradiation. Each emission, absorption, and scattering event is treated probabilistically and self-consistently based on the local energy spectrum and electromagnetic environment, allowing the simulation to resolve transient charge fluctuations and spatial heterogeneities within regolith micro-cavities.

\section{Incorporation of Charge Dissipation}\label{app:charge-dissipation}

While the Geant4 framework accounts for the stochastic deposition of charge from incident currents, it does not inherently model the subsequent charge relaxation, or dissipation, through the dielectric material. The net temporal change in surface charge density, $\dot{\sigma}$, is governed by the continuity equation, given as \cite{Zimmerman_2016}:

\begin{equation}
\dot{\sigma}
=
j\exp\left(-\frac{\sigma}{\Sigma}\right)
-
\frac{\vartheta}{\epsilon_r\epsilon_0}\sigma,
\label{eq:charge_diff}
\end{equation} 
where the first term represents the net charging current (e.g., from PE, secondary emission, SW flux), and the second term is the ohmic dissipation term, which must be incorporated explicitly to model charge transport within the material. For lunar regolith, the temperature-dependent electrical conductivity, $\vartheta(T)$, is \cite{Grard1973}:

\begin{equation}
\vartheta(T)
=
6\times10^{-18}\exp(0.0230T)\ \SI{}{\ohm^{-1}m^{-1}}.
\label{eq:conductivity}
\end{equation} 
A constant electrical conductivity of $\vartheta\sim10^{-13}\ \Omega^{-1}\text{m}^{-1}$ (corresponding to an equivalent temperature of $\SI{425}{K}$) is used for all results presented in this paper. Over a simulation time of $\SI{10}{s}$, the induced temperature changes based on local charge accumulation are negligible ($<\SI{2}{K}$); therefore, the time dependence of the dissipation term can be neglected. However, this effect can be incorporated for systems in which temperature variations relative to the ambient thermal bath are significant.

To implement charge dissipation, the discrete charges deposited by Geant4 are first assigned to the corresponding voxels of the charge octree, $O_{\mathrm{charge}}$. For voxels associated with the grain surface, the net deposited charge is converted to a surface charge density, $\sigma$, using the grain-surface area represented within that voxel. Thus, the three-dimensional voxel serves as a computational region for grouping the discrete deposited charges, while $\sigma$ represents the charge density associated with the grain surface rather than a volumetric charge density in the limit the penetration depth goes to zero. At each iteration, the ohmic dissipation term in Eq.~\ref{eq:charge_diff} is then applied to reduce the accumulated surface charge according to the material conductivity, dielectric constant, and elapsed simulation time to macroscopically leak the charge. This treatment represents local charge relaxation through the weakly conducting dielectric rather than explicitly tracking the microscopic transport of charge between neighboring voxels. Charge removed through this dissipation term is therefore not transferred to another surface voxel, but represents charge conducted away from the local surface region through the dielectric material. The resulting charge distribution is saved to the master charge list and used to calculate the electric field for the subsequent simulation.

\section{Simulation Runtimes on HPC}\label{app:HPC-runtimes}

\begin{figure}[!ht]
\centering
\includegraphics[width=1\textwidth]{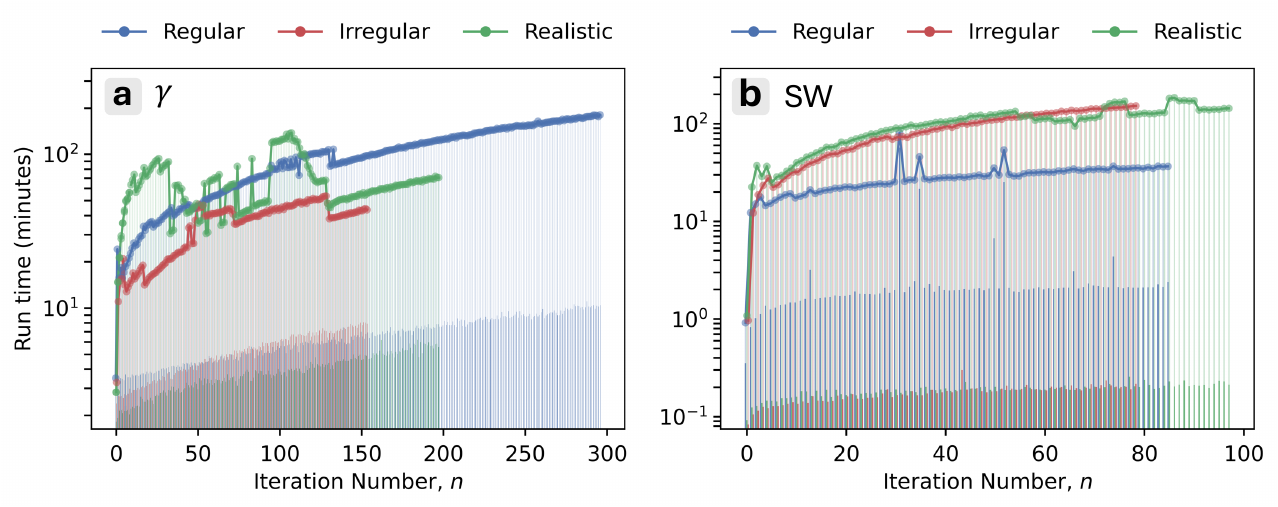}
\caption{Wall-clock runtimes per iteration for the regular (blue), irregular (red), and realistic (green) geometries on the HPC cluster. The darker portion of each bar shows the particle-tracking runtime, the lighter portion shows the electric field precomputation runtime, and the overlaid marker indicates the total runtime. Panel (a) shows the photon case, and panel (b) shows the SW case. Computational resources are described in Section~\ref{sec:hpc-resources}.}
\label{fig:runtimes}
\end{figure}

Our code was executed on the PACE HPC cluster at Georgia Tech using SLURM-managed batch jobs. Each iteration was run on a single node with 24 CPU threads and required at least 8 GB of memory per process (see Section~\ref{sec:hpc-resources}), so the wall-time depended strongly on the problem size and the resulting octree complexity. In practice, the runtime was governed primarily by thread interdependence when parallelizing octree generation and input/output write speeds, rather than by the physics calculations themselves. Figure~\ref{fig:runtimes} summarizes the runtimes per iteration for the regular, irregular, and realistic geometries in the SW and photon cases. In Fig.~\ref{fig:runtimes}a,b, the darker segment of each bar indicates the particle-tracking time, while the lighter segment indicates the precomputation time; the total runtime is shown by the overlaid marker and connecting line. For both the cases, the total runtimes are dominated by precomputation. The SW runs are heavily dominated by precomputation (lighter segments), with radiation tracking contributing $<1\%$ of the total runtime; the exception is the regular geometry, for which the larger world size and, thus, proportionally smaller hadronic tracking cuts increase the particle-tracking contribution to approximately $10\%$ of the total runtime. In contrast, the photon runs have a larger particle-tracking contribution (darker segments) because $3-125\times$ more particles are tracked per iteration (Table~\ref{tab:simulation-parameters}). The particle-tracking accounts for approximately $6-14\%$ of the total runtime for the photon case. Future work will focus on improving computational efficiency in both components of the simulation. For the field precomputation, we will investigate a shared-memory approach within HPC cores to reduce computational redundancies. For the Geant4 radiation-transport calculations, we intend to explore parallelizing the radiation transport using Geant4's built-in multi-threading capabilities.

%\end{linenumbers}

\bibliographystyle{elsarticle-num} 
\bibliography{references}

\end{document}